\documentclass[12pt,prd,tightenlines,nofootinbib]{revtex4}
\usepackage{bm}
\usepackage{graphics}
\usepackage{rotating}
\usepackage{epsfig}
\usepackage{slashed,amsmath,float}
\usepackage{mathrsfs}%花写体小花\mathscr{P}
\usepackage{amsfonts}%花写体大花\mathfrak{P}
\usepackage{graphicx,subfigure}
\usepackage{color}
\usepackage{ulem,bbm}
\usepackage{diagbox}
\usepackage{makecell}

\begin{document}

\title{
Forward-backward asymmetries in $\Lambda_b \rightarrow \Lambda l^+ l^-$  in Bethe-Salpeter equation approach}
\author{Liang-Liang Liu $^{a}$}
\email{corresponding  author. liu06_04@sxnu.edu.cn}
\author{Su-Jun Cui $^{a}$}
\author{Jing Xu $^{b}$}
\author{ Xin-Heng Guo $^{c}$}
\email{ corresponding  author. xhguo@bnu.edu.cn}

\affiliation{\footnotesize (a)~College of Physics and information engineering, Shanxi Normal University, Taiyuan 030031, People's Republic of China}

\affiliation{\footnotesize (b)~Department of Physics, Yan-Tai University, Yantai 264005, People's Republic of China}

\affiliation{\footnotesize (c)~College of Nuclear Science and Technology, Beijing Normal University, Beijing 100875, People's Republic of China}
\begin{abstract}
Using the Bethe-Salpeter equation (BSE) we investigate the forward-backward asymmetries $( A _{FB}) $ in $\Lambda_b \rightarrow \Lambda l^+ l^-(l=e,\mu,\tau)$ in the quark-diquark model. This approach provides precise form factors that are different from those of QCD sum rules.
We calculate the rare decay form factors for $\Lambda_b \rightarrow \Lambda l^+ l^-$ and investigate the (integrated) forward-backward asymmetries in these decay channels.
We find that the integrated $A^l_{FB}$, $\bar{A}^l_{FB}(\Lambda_b \rightarrow \Lambda e^+ e^-) \simeq -0.1371 $, $\bar{A}^l_{FB}(\Lambda_b \rightarrow \Lambda  \mu^+ \mu^- ) \simeq -0.1376 $, $\bar{A}^l_{FB}(\Lambda_b \rightarrow \Lambda \tau^+ \tau^-) \simeq -0.1053 $, the hadron side asymmetries $\bar{A}^h_{FB}(\Lambda_b \rightarrow \Lambda  \mu^+ \mu^-)\simeq -0.2315$, the lepton-hadron side asymmetries $\bar{A}^{lh}_{FB}(\Lambda_b \rightarrow \Lambda  \mu^+ \mu^-)\simeq 0.0827$, the longitudinal polarization fractions $\bar{F}_L(\Lambda_b \rightarrow \Lambda  \mu^+ \mu^-)\simeq 0.5681$.
\end{abstract}

\maketitle

\section{Introduction}

Decays of hadrons involving the flavor changing neutral current (FCNC) transition such as $\Lambda_b \rightarrow \Lambda l^+ l^-$ can give essential information about the inner structure of hadrons, reveal the nature of the electroweak interaction, and provide model-independent information about physical quantities such as Cabibbo-Kobayashi-Maskawa (CKM) matrix elements.
The rare decay $\Lambda_b \rightarrow \Lambda \mu^+ \mu^- $ was first observed by the CDF Collaboration in 2011 \cite{PRL107-201802}.
Some experimental progresses about  $\Lambda_b \rightarrow \Lambda l^+ l^-$ were also achieved \cite{ PLB725-25, PRL123-031801, JHEP09-146, JHEP06-115} and the radiative decay $\Lambda_b \rightarrow \Lambda \gamma$ was observed in 2019 \cite{PRL123-031801} by LHCb.
The LHCb Collaboration determined the forward-backward asymmetries ($A^l_{FB}$) of the decay $\Lambda_b \rightarrow \Lambda \mu^+ \mu^-$ to be $ {A}^l_{FB}(\Lambda_b \rightarrow \Lambda \mu^+ \mu^-)=-0.05\pm0.09$~(stat) $\pm0.03$~(syst), $ {A}^h_{FB}(\Lambda_b \rightarrow \Lambda \mu^+ \mu^-)=-0.29\pm0.09$~(stat) $\pm0.03$~(syst), $ F_L(\Lambda_b \rightarrow \Lambda \mu^+ \mu^-)=0.61^{+0.11}_{-0.14} \pm 0.03$~(syst) at the low dimuon invariant mass squared range $15 < q^2<20$ GeV$^2$ in 2015 \cite{JHEP06-115}.
However, these number were updated in 2018 to be $\bar{A}^l_{FB}(\Lambda_b \rightarrow \Lambda \mu^+ \mu^-)=-0.39\pm0.04$ (stat) $\pm0.01$ (syst), $ {A}^h_{FB}(\Lambda_b \rightarrow \Lambda \mu^+ \mu^-)=-0.3\pm0.05$~(stat) $\pm0.02$~(syst), $\bar{A}^{lh}_{FB}(\Lambda_b \rightarrow \Lambda \mu^+ \mu^-)=0.25\pm0.04$ (stat) $\pm0.01$ (syst), in the same invariant mass squared region \cite{JHEP09-146}.
It is noted that the $A^l_{FB}$ is much lager than the previous one.
In this work, we will study the $ A _{FB}$ of $\Lambda_b \rightarrow \Lambda l^+ l^-$ in Bethe-Salpeter equation (BSE) approach.
Theoretically, there are a few works on the study of  $ {A}_{FB}( \Lambda_b\rightarrow \Lambda l^+ l^-)$ \cite{PRD87-074031, PRD64-074001, PLB516-327, JHEP12-067, NPB709-115, IJMPA27-1250016, PRD96-053006,EPJC59-861,JHEP01-155,JHEP02-179,IJMPA30-1550172,PRD103-013007}.
Ref. \cite{PRD64-074001} (\cite{PLB516-327}) gave the integrated forward-backward asymmetries $ \bar{A}^l_{FB}( \Lambda_b\rightarrow \Lambda \mu^+ \mu^-)=-0.13$ ($-0.12$) and $ \bar{A}_{FB}( \Lambda_b\rightarrow \Lambda \tau^+ \tau^-)=-0.04$ ($-0.03$), while the results of Ref. \cite{PRD87-074031} were $ \bar{A}^l_{FB}( \Lambda_b\rightarrow \Lambda e^+ e^-)=1.2 \times 10^{-8}$, $ \bar{A}^l_{FB}( \Lambda_b\rightarrow \Lambda \mu^+ \mu^-)= 8 \times 10^{-4}$ and $ \bar{A}^l_{FB}( \Lambda_b\rightarrow \Lambda \tau^+ \tau^-)=9.6\times 10^{-4}$.
Ref. \cite{JHEP12-067} analyzed the differential $\bar{A}_{FB}(\Lambda_b\rightarrow \Lambda l^+ l^-)$ in the heavy quark limit.
Using the nonrelativistic quark model, Ref. \cite{IJMPA27-1250016} investigated the lepton-side forward-backward asymmetries $\bar{A}^l_{FB}(\Lambda_b\rightarrow \Lambda l^+ l^-)$.
In the quark-diquark model, Ref. \cite{PRD96-053006} investigated the lepton-side forward-backward asymmetries $A_{FB}$, the hadron-side forward-backward asymmetries $A_{FB}^h$ and the hadron-lepton forward-backward asymmetries $A^{hl}_{FB}$.
In approach of the light-cone sum rules, Refs. \cite{EPJC59-861, JHEP02-179} investigated the rare decays of $\Lambda_b \rightarrow \Lambda \gamma$ and  $\Lambda_b \rightarrow \Lambda l^+ l^-$.
Ref. \cite{JHEP01-155} investigated the phenomenological potential of the rare decay $\Lambda_b \rightarrow \Lambda l^+ l^- $ with a subsequent, self-analyzing $\Lambda_b \rightarrow N \pi $ transition.
With the FFs extracted from a constituent quark model, Ref. \cite{IJMPA30-1550172} investigated the rare weak dileptonic decays of the $\Lambda_b$ baryon.
Ref. \cite{PRD103-013007} studied ${\cal B}_1 \rightarrow {\cal B}_2 l^+ l^-$ (${\cal B}_{1,2}$ are spin $1/2$ baryons) with the SU(3) flavor symmetry.
The Form Factors (FFs) of $\Lambda_b \rightarrow \Lambda$ are different in different models.
Generally, the number of independent FFs of $\Lambda_b \rightarrow \Lambda$ can be reduced to 2 when working in heavy quark limit \cite{NPB355-38},
\begin{eqnarray} \label{FFs-HQET}
\langle \Lambda(p)| \bar{s} \Gamma b | \Lambda_b(v) \rangle = \bar{u}_{\Lambda} (F_1(q^2) +F_2 (q^2) \slashed{v} )\Gamma u_{\Lambda_b}(v),
\end{eqnarray}
where $\Gamma= \gamma_\mu,~ \gamma_\mu \gamma_5,~q^\nu \sigma_{\nu\mu}$, and $q^\nu \sigma_{\nu\mu}\gamma_5$, $q^2$ is the transformed momentum squared.
The FFs ratio $R(q^2)= F_2(q^2)/F_1(q^2)$  was regarded as a constant in many works assuming the same shape for $F_1$ and $F_2$ which was derived from QCD sum rules in the framework  of the heavy quark effective theory \cite{PRD64-074001}.
For example, in Ref. \cite{PRD64-074001,PLB516-327} the $q^2$ dependence of FFs $F_i~(i=1,2)$ are give as follows:
\begin{eqnarray}
  F_i(q^2) &=& \frac{F_i(0)}{1-a q^2+b q^4},
\end{eqnarray}
where $a$ and $b$ are constants. Using experimental data for the semileptonic decay $\Lambda_c \rightarrow \Lambda e^+ \nu_e$ ($m^2_{\Lambda} \leq q^2 \leq m^2_{\Lambda_c} $), the CLEO Collaboration gave the ratio $R=-0.35\pm0.04$ (stat) $\pm0.04$ (syst) \cite{PRL94-191801}.
In Ref. \cite{JPG24-979} the authors investigated $\Lambda_b \rightarrow \Lambda \gamma $ giving $R=-0.25\pm0.14\pm0.08$.
In Refs. \cite{PRD63-114024,PRD64-074001, PLB516-327} the authors investigated the baryonic decay $\Lambda_b \rightarrow \Lambda l^+ l^- $ and obtained $R=-0.25$.
In Ref. \cite{NPB649-168} the relation $F_2(q^2)/F_1(q^2) \approx F_2(0)/F_1(0) $ was given.
However, according to the pQCD scaling law \cite{PRD22-2157,PRD11-1309, PPNP59-694}, the FFs should not have the same shape.
Using Stech's approach in Ref. \cite{PRD53-4946} the authors obtained the FFs ratio $R(q^2) \propto -1/q^2 $ .
From the data in Ref. \cite{PRD59-114022}, we can estimate the value of $R$ and find it changes from $-0.83$ to $-0.32$ which is not a constant.
In our previous works \cite{CPC44-083107, EPJC80-193}, we found that the ratio $R$ is not a constant in the $\Lambda_b$ rare decay in a large momentum region where we did not consider the long distance contributions because they have a small effect on FFs of this decay \cite{PLB367-362,PRD51-1215}.
In these works $\Lambda_b$ ($\Lambda$) is regarded as a bound state of two particles: a quark and a scalar diquark. This model has been used to study many heavy baryons \cite{PRD54-4629}.
Using the kernel of the BSE including a scalar confinement term and a one-gluon-exchange terms and the covariant instantaneous approximation, we obtianed the Bethe-Salpeter (BS) wave functions of $\Lambda_b$ and $\Lambda$ \cite{CPC44-083107, EPJC80-193}.
In the present work, we will recalculate the FFs of $\Lambda_b \rightarrow \Lambda$ in this model.

This paper is organized as follows. In Section II, we will derive the general FFs and $A_{FB}$ for $ \Lambda_b \rightarrow \Lambda l^+ l^-$ in the BS equation approach.
In Section III the numerical results for $A_{FB}$ and $\bar{A}_{FB}$ of $\Lambda_b \rightarrow \Lambda l^+ l^- $ will be given.
Finally, the summary and discussion will be given in Section V.

\section{THEORETICAL FORMALISM}\label{sec2}

\subsection{The BSE for $\Lambda_b(\Lambda)$}

\begin{figure}[!htb]
\begin{center}
\includegraphics[width=7.5cm]{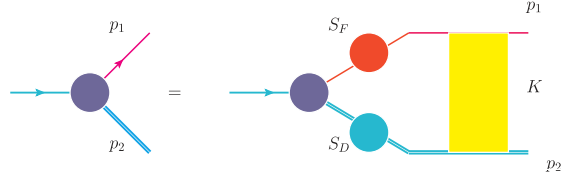}
\caption{The BS equation for $\Lambda_b(\Lambda)$ in momentum space (K is the interaction kernel)}\label{BSE}
\end{center}
\end{figure}

As shown in Fig. \ref{BSE}, following our previous work the BSE of $\Lambda_b(\Lambda)$ in momentum space satisfies the integral equation \cite{CPC42-103106, PRD95-054001, PRD87-076013, PRD91-016006, PLB954-97, PRD86-056006, PRD76-056004,EPJC80-193,CPC44-083107}
\begin{eqnarray}\label{chi-p}
\chi_P(p) =  S_F(\lambda_1 P+p)\int \frac{d^4 q}{(2 \pi)^4}  K (P,p,q)\chi_P(q)S_D(\lambda_2 P-p),
\end{eqnarray}
where $K(P,p,q)$ is the kernel which is defined as the sum of the two particles irreducible diagrams, $S_F$ and $S_D$ are the propagators of the quark and the scalar diquark, respectively.
$\lambda_{1(2)}=m_{q(D)}/(m_q+m_D)$, with $m_{q(D)}$ being the mass of quark (diquark) and
$P$ is the momentum of the baryon.

We assume the kernel has the following form:
\begin{eqnarray}
- i K(P,p,q) &=&   I\otimes I V_1(p,q)+ \gamma_\mu \otimes (p_2+q_2)^\mu  V_2(p,q),
\end{eqnarray}
where $V_1$ arises from the scalar confinement and $V_2$ is from the one-gluon-exchange diagram.
According to the potential model, $V_1$ and $V_2$ have the following forms in the covariant instantaneous approximation ($p_l=q_l$) \cite{CPC44-083107,PRD87-076013, PRD86-056006, PRD76-056004, EPJC80-193}:
\begin{eqnarray}\label{VV}
  \tilde{V}_1(p_t-q_t) &=& \frac{8 \pi \kappa}{[(p_t-q_t)^2+\mu^2]^2} - (2\pi)^2\delta^3(p_t-q_t)\int \frac{d^3k}{(2\pi)^3} \frac{8 \pi \kappa}{(k^2+\mu^2)^2}, \\
  \tilde{V_2} (p_t-q_t)&=&- \frac{16 \pi }{3}\frac{\alpha^2_{seff} Q^2_0}{[(p_t-q_t)^2+\mu^2][(p_t-q_t)^2+Q_0^2]},
\end{eqnarray}
where $ \mu $ is a small parameter to avoid the divergence in numerical calculations this parameter is taken to be small enough so that the results are not sensitive to it, the parameters $\kappa$ and $\alpha_{seff}$ are related to scalar confinement and the one-gluon-exchange diagram, respectively.
$q_t$ is the transverse projection of the relative momentum along the momentum $P$ which is defined as $p_l = \lambda_1 P -v\cdot p,~p_t^\mu = p^\mu-(v\cdot p)p^\mu$ ($v^\mu=P^\mu/M$), $ q_t^\mu = q^\mu -(v \cdot q) v^\mu$, and $q_l=\lambda_2 P -v \cdot q$.
The second term of $\tilde{V}_1$ is introduced to avoid infrared divergence at the point $ p_t=q_t$, and $\mu$ is a small parameter to avoid the divergence in numerical calculates.
Analyzing the electromagnetic FFs of the proton, it was found that $Q_0^2=3.2$ GeV$^2$ can lead to consistent results with the experimental data \cite{ZPA347-109}.

The propagators of the quark and the diquark can be written as the following:
\begin{eqnarray}\label{SF}
  S_F(p_1) = i \slashed{v} \bigg[ \frac{\Lambda_q^+ }{ M -p_l -\omega_q  +i \epsilon} +\frac{\Lambda_q ^-}{ M -p_l +\omega  -i \epsilon}\bigg],
\end{eqnarray}
\begin{eqnarray}\label{SD}
  S_D(p_2)  = \frac{i}{2 \omega_D} \bigg[\frac{1}{ p_l-\omega_D+i \epsilon} -\frac{1}{ p_l+ \omega_D-i\epsilon}\bigg],
\end{eqnarray}
where $\omega_q = \sqrt{m^2-p_t^2}~\text{and}~\omega_D = \sqrt{m_D^2-p_t^2} $, $M$ is the mass of the baryon, $\Lambda^\pm$ are the projection operators which are defined as
\begin{eqnarray}
2 \omega_q \Lambda^\pm_q&=& \omega_q \pm  \slashed{v}(\slashed{p}_t+m) ,
\end{eqnarray}
and satisfy the following relations:
\begin{eqnarray}
\Lambda_q^\pm \Lambda_q^\pm &=& \Lambda^\pm_q,~\Lambda^\pm_q \Lambda^\mp_q  =  0.
\end{eqnarray}
Generally, we need two scalar functions to describe the BS wave function of $\Lambda_b(\Lambda)$ \cite{CPC42-103106, PRD95-054001, PRD91-016006},
\begin{eqnarray}
  \chi_P(p) &=& (f_1(p_t^2)+\slashed{p}_t f_2(p_t^2))u(P), \label{wf-ls}
\end{eqnarray}
where $f_i, (i=1,2)$ are the Lorentz-scalar functions of $p_t^2$, and $u(P)$ is the spinor of baryon.

Defining $\tilde{f}_{1(2)}=\int \frac{d p_l}{2 \pi}f_{1(2)}$, and using the covariant instantaneous approximation, the scalar BS wave functions satisfy the following coupled integral equations:

\begin{eqnarray}\label{BS:f12}
&& \tilde{f}_1(p_t) =\int \frac{d^3q_t}{(2\pi)^3} M_{11}(p_t,q_t) \tilde{f}_1(q_t)+  M_{12}(p_t,q_t) \tilde{f}_2(q_t), \\
 &&\tilde{f}_2(p_t) = \int \frac{d^3q_t}{(2\pi)^3}  M_{21}(p_t,q_t) \tilde{f}_1(q_t) +  M_{22}(p_t,q_t) \tilde{f}_2(q_t),
\end{eqnarray}
where
\begin{eqnarray}
M_{11}(p_t,q_t)=\frac{(\omega_q  +m ) (\tilde{V}_1+ 2 \omega_D \tilde{V}_2)-   p _t \cdot ( p _t+ q _t) \tilde{V}_2}{4 \omega_D \omega_q(-M + \omega_D+ \omega_q)} - \nonumber\\ \frac{(\omega_q -m )(\tilde{V}_1- 2\omega_D \tilde{V}_2)+   p _t\cdot( p _t+ q _t)  \tilde{V}_2}{4 \omega_D \omega_c(M + \omega_D+ \omega_q)},
\end{eqnarray}

\begin{eqnarray}
M_{12}(p_t,q_t)=\frac{-  (\omega_q+m ) ( q _t + p _t)\cdot q_t\tilde{V}_2 +  p _t\cdot q_t(\tilde{V}_1- 2 \omega_D \tilde{V}_2)}{4 \omega_D \omega_c(-M + \omega_D+ \omega_c)}- \nonumber\\ \frac{(m - \omega_q )  ( q _t + p _t)\cdot q _t \tilde{V}_2 -   p _t\cdot q _t  (\tilde{V}_1+ 2\omega_D \tilde{V}_2)}{4 \omega_D \omega_q(M + \omega_D+ \omega_q)},
\end{eqnarray}

\begin{eqnarray}
M_{21}(p_t,q_t)= \frac{(\tilde{V}_1+ 2 \omega_D \tilde{V}_2)-( -\omega_q+m ) (1+\frac{ q _t \cdot p _t }{ p^2_t })\tilde{V}_2}{4 \omega_D \omega_q(-M + \omega_D+ \omega_q)}    - \nonumber\\
\frac{- (\tilde{V}_1- 2\omega_D \tilde{V}_2)+(\omega_q  + m )(1+\frac{ q _t \cdot p _t }{ p^2_t }) \tilde{V}_2)}{4 \omega_D \omega_q(M + \omega_D+ \omega_q)},
\end{eqnarray}

\begin{eqnarray}
M_{22}(p_t,q_t)=  \frac{(m  -\omega_q)( \tilde{V}_1+ 2  \omega_D \tilde{V}_2)) p_t \cdot q_t - p^2_t ( q^2_t+  p_t \cdot q_t) \tilde{V}_2}{4  p^2_t \omega_D \omega_q(-M + \omega_D+ \omega_q)} - \nonumber\\
\frac{ (m +\omega_q) (-\tilde{V}_1- 2 \omega_D \tilde{V}_2))  p_t \cdot q_t +  p^2_t (  q^2_t+  p_t \cdot q_t)\tilde{V}_2 }{4  p^2_t \omega_D \omega_q(M + \omega_D+ \omega_q)}.
\end{eqnarray}

When the mass of the $b$ quark goes to infinity \cite{PRD54-4629}, the propagator of the $b$ quark satisfies the relation $\slashed{v} S_F(p_1)= S_F(p_1)$ and can be reduced to
\begin{eqnarray}\label{SF-HQ}
  S_F(p_1) = i \frac{ 1+ \slashed{v}  }{  2 (E_0+m_D -p_l+ i \epsilon) },
\end{eqnarray}
where $E_0=M-m-m_D$ is the binding energy.
Then, the BS wave function of $\Lambda_b$ has the form $\chi_P (v)=\phi (p)u_{\Lambda_b}(v,s)$, with $\phi(p)$ being the scalar BS wave function \cite{PRD54-4629}, and the BS equation for $\Lambda_b$ can be replaced by
\begin{eqnarray}\label{BS:hq}
  \phi(p) &=& -\frac{i}{(E_0+m_D-p_l+i \epsilon)( p_l ^2-\omega^2_D)}\int \frac{d^4 q }{(2\pi)^4}(\tilde{V}_1+2  p_l \tilde{V}_2)\phi(q).
\end{eqnarray}

Generally, one can take $E_0$ to be about $-0.14$ GeV and $\kappa$ to be about $0.05$ GeV$^3$ \cite{CPC44-083107, EPJC80-193}.

\subsection{The Asymmetries of $ \Lambda_b \rightarrow \Lambda l^+ l^-  $  decays}

In the SM, the $\Lambda_b\rightarrow \Lambda l^+l^-$ ($l=e,\mu, \tau$) transitions are described by the $b\rightarrow s l^+l^-$ at the quark level.
The Hamiltonian for the decay of $b\rightarrow s l^+l^-$ is given by

\begin{eqnarray}
\mathcal{H}( b\rightarrow s l^+l^-) &=&  \frac{G_F\alpha}{2 \sqrt{2}\pi}V_{tb}V^*_{ts}\bigg[ C^{eff}_9 \bar{s} \gamma_{\mu}(1-\gamma_5)b \bar{l}\gamma^{\mu}l  - i C^{eff}_{7}\bar{s}\frac{2 m_b\sigma_{\mu\nu} q^{\nu}}{q^2}(1+\gamma_5) b \bar{l}\gamma^{\mu}l  \nonumber \\
~~~~~~~~~~&+&C_{10} \bar{s}\gamma_{\mu}(1-\gamma_5) b \bar{l}\gamma^{\mu}\gamma_5l  \bigg],
\end{eqnarray}
where $G_F$ is the Fermi coupling constant, $\alpha$ is the fine structure constant at the Z mass scale, $V_{ts}$ and $V_{tb}$ are the CKM matrix elements, $q$ is the total momentum of the lepton pair and $C_i~(i=7,~9,~10)$ are the Wilson coefficients.
$C^{eff}_7=-0.313$, $C^{eff}_9=4.334$, $C_{10}=-4.669$ \cite{JHEP10-118, PRD79-074007, EPJC40-565}.
The relevant matrix elements can be parameterized in terms of the FFs as follows:
\begin{eqnarray}\label{FFs}
 \langle \Lambda(P^\prime) | \bar{s}\gamma_{\mu}b | \Lambda_b(P)\rangle &=& \bar{u}_{\Lambda}(P^\prime)(g_1\gamma^\mu+ ig_2\sigma^{\mu\nu}q_{\nu}+g_3q_\mu)u_{\Lambda_b}(P),\nonumber\\
 \langle \Lambda(P^\prime) | \bar{s}\gamma_{\mu}\gamma_{5}b  | \Lambda_b(P)\rangle &= & \bar{u}_{\Lambda}(P^\prime)(t_1\gamma^\mu+it_2\sigma^{\mu\nu}q_{\nu}+t_3q^\mu)\gamma_5u_{\Lambda_b}(P),\nonumber\\
 \langle \Lambda (P^\prime) | \bar{s}i\sigma^{\mu\nu}q^{\nu}b | \Lambda_b(P)\rangle &= & \bar{u}_{\Lambda}(P^\prime)(s_1\gamma^\mu+is_2\sigma^{\mu\nu}q_{\nu}+s_3q^\mu)u_{\Lambda_b}(P),\nonumber\\
 \langle \Lambda (P^\prime)  |  \bar{s}i\sigma^{\mu\nu}\gamma_5q^{\nu}b | \Lambda_b(P)\rangle &= & \bar{u}_{\Lambda}(P^\prime)(d_1\gamma^\mu+id_2\sigma^{\mu\nu}q_{\nu}+d_3q^\mu)\gamma_5u_{\Lambda_b}(P),
\end{eqnarray}
 where $P (P^\prime)$ is the momentum of the $\Lambda_b$($\Lambda$), $q^2=(P-P^\prime)^2$ is the transformed momentum squared, $g_i$, $t_i$, $s_i $, $d_i $ ($i=1,2$ and 3) are the transition FFs which are Lorentz scalar functions of $q^2$.
The $\Lambda_b$ and $\Lambda$ states can be normalized as the following:
\begin{eqnarray}
  \langle \Lambda(P^\prime)|\Lambda(P)\rangle &=& 2 E_\Lambda (2\pi)^3 \delta^3(P-P^\prime), \\
  \langle \Lambda_b(v^\prime,P^\prime)|\Lambda_b(v,P)\rangle &=& 2 v_0(2\pi)^3 \delta^3(P-P^\prime).
\end{eqnarray}
 Comparing Eq. (\ref{FFs-HQET}) with Eq. (\ref{FFs}) we obtain the following relations:
\begin{eqnarray}
 & & g_1~=~t_1~=~s_2~=~d_2~=~\bigg(F_1+\sqrt{r}F_2\bigg),\nonumber\\
 & & g_2~=~t_2~=g_3~=~t_3~=~\frac{1}{m_{\Lambda_{b}}}F_2, \nonumber\\
 & & s_3~=~  F_2 (\sqrt{r}-1),~ d_3~=~ F_2(\sqrt{r}+1), \nonumber\\
 & & s_1 ~=~ d_1~=~ F_2 m_{\Lambda_b}  (1+r-2\sqrt{r}\omega),
\end{eqnarray}
where $r=m_\Lambda^2/m_{\Lambda_b}^2$ and $\omega= (M_{\Lambda_b}^2+M_{\Lambda}^2-q^2)/(2M_{\Lambda_b} M_{\Lambda})=v\cdot P^\prime/m_{\Lambda}$.
The transition matrix for $\Lambda_b\rightarrow \Lambda$ can be expressed in terms of the BS wave functions of $\Lambda_b$ and $\Lambda$,
\begin{eqnarray}\label{FFs-BS}
  \langle \Lambda (P^\prime)|\bar{d}\Gamma b|\Lambda_b(P)\rangle =\int\frac{d^4p}{(2\pi)^4} \bar{\chi}_{P^\prime}(v^\prime)\Gamma \chi_P(p)S^{-1}_D(p_2).
\end{eqnarray}

When $\omega \neq 1$, one can obtain the following expression by taking Eq. (\ref{wf-ls}) and (\ref{BS:hq}) into Eq. (\ref{FFs-BS}):

\begin{eqnarray}
  F_1 &=& k_1- \omega k_2, \\
    F_2 &=& k_2,
\end{eqnarray}
where
\begin{eqnarray}
  k_1(\omega)&=&\int \frac{d^4p}{(2 \pi)^4} f_1(p^\prime) \phi(p) S^{-1}_D(p_2), \\
  k_2(\omega) &=& \frac{1}{1-\omega^2} \int \frac{d^4 p}{(2\pi)^4} f_2(p^\prime) p^\prime_t \cdot v \phi(p) S^{-1}_D.
\end{eqnarray}

The decay amplitude of $\Lambda_b \rightarrow \Lambda l^+ l^-$ can be rewritten as the following:
\begin{eqnarray}
 \mathcal{M}(\Lambda_b\rightarrow \Lambda l^{+} l^{-})&=&\frac{G_F \lambda_t}{2\sqrt{2}\pi}  \big[\bar{l}\gamma_{\mu}l\{\bar{u}_{\Lambda}[\gamma_{\mu}(A_1+B_1+ (A_1-B_1)\gamma_5 ) \nonumber\\
 & +& i\sigma^{\mu\nu}p_{\nu}(A_2+B_2+ (A_2-B_2)\gamma_5 )]u_{\Lambda_b}\} \nonumber\\
&+&\bar{l}\gamma_{\mu}\gamma_5l\{\bar{u}_{\Lambda}[\gamma^{\mu}(D_1+E_1+ (D_1-E_1)\gamma_5 ) \nonumber \\
 &+&i\sigma^{\mu\nu}p_{\nu}(D_2+E_2+ (D_2-E_2)\gamma_5 )\nonumber\\
 &+&p^{\mu}(D_3+E_3+ (D_3-E_3)\gamma_5 )]u_{\Lambda_b}\}\big],
\end{eqnarray}
where $A_i$, $B_i$ and $D_j$, $E_j$ ($i=1,2$ and $j=1,2,3$) are defined as the following:
\begin{eqnarray}
&&A_i=\frac{1}{2}\bigg\{C^{eff}_{9}(g_i-t_i)-\frac{2C^{eff}_7 m_b}{q^2}(d_i +s_i )\bigg\},\nonumber\\
& &B_i = \frac{1}{2}\bigg\{C^{eff}_{9}(g_i+t_i) - \frac{2C^{eff}_7m_b}{q^2}(d_i -s_i )\bigg\}, \nonumber\\
& &D_j = \frac{1}{2}C_{10}(g_j-t_j), ~E_j=\frac{1}{2}C_{10}(g_j+t_j).
\end{eqnarray}

\begin{figure*}[btp]
\centering
\includegraphics[width=7cm]{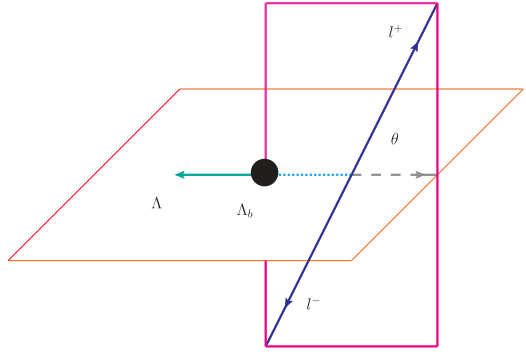}
\caption{Definition of the angle $\theta$ in the decay $\Lambda_b \rightarrow \Lambda l^- l^+$.}\label{Fig:Rare}
\end{figure*}

In the physical region ( $ \omega = ( m_{\Lambda_b}^2 + m_{\Lambda}^2 -q^2 )/(2m_{\Lambda_b}m_{\Lambda})$), the decay rate of $\Lambda_b\rightarrow \Lambda l^+l^-$ is obtained as the following:

\begin{eqnarray}
\frac{d\Gamma(\Lambda_b\rightarrow \Lambda l^+l^-)}{d\omega d \cos \theta}=\frac{G^2_F\alpha^2}{2^{14}\pi^5m_{\Lambda_b}} |V_{tb}V^*_{ts}|^2v_l\sqrt{\lambda(1,r,s)} \mathcal{M}(\omega, \theta)  ,
\end{eqnarray}
where  $s=   1 +r  - 2 \sqrt{r} \omega $, $ \lambda(1,r,s)=1+r^2+s^2-2r-2s-2rs$,  $v_l=\sqrt{1-\frac{4m^2_l}{s m^2_{\Lambda_b}}}$, and the decay amplitude is given as the following \cite{EPJC45-151}:

\begin{eqnarray}
   \mathcal{M}(\omega,\theta) &=& \mathcal{M}_0(\omega) +\mathcal{M}_1(\omega) \cos \theta +\mathcal{M}_2(\omega)  \cos^2 \theta,
\end{eqnarray}
where $\theta $ is the polar angle, as is shown in Fig. \ref{Fig:Rare}.

\begin{eqnarray}
  \mathcal{M}_0(\omega)&&=32m^2_l m^4_{\Lambda_b}s(1+r-s)(|D_3|^2+|E_3|^2) \nonumber\\
  &&+64m^2_lm^3_{\Lambda_b}(1-r-s)Re(D^*_1E_3+D_3E^*_1)\nonumber\\
& &+64m^2_{\Lambda_b}\sqrt{r}(6m^2_l-M^2_{\Lambda_b}s)Re(D_1^*E_1)\nonumber\\
&& + {64m^2_lm^3_{\Lambda_b}\sqrt{r}\big(2m_{\Lambda_b}s Re(D^*_3E_3) +(1-r+s)Re(D^*_1D_3+E^*_1E_3)\big) }\nonumber\\
&&+32m^2_{\Lambda_b}(2m^2_l+m^2_{\Lambda_b}s)\bigg\{(1-r+s)m_{\Lambda_b}\sqrt{r}Re(A^*_1A_2+B^*_1B_2)\nonumber\\
& &-m_{\Lambda_b}(1-r-s)Re(A^*_1B_2+A^*_2B_1)  -2\sqrt{r}\big(Re(A^*_1B_1)+m^2_{\Lambda_b}s Re(A^*_2B_2)\big) \bigg \}\nonumber\\
& &+ 8 m^2_{\Lambda_b}\bigg[4m^2_l(1-r-s)+m^2_{\Lambda_b}((1+r)^2- s^2)\bigg](|A_1|^2+|B_1|^2)\nonumber\\
&&+8m^4_{\Lambda_b}\bigg\{4m^2_l[\lambda+(1+r-s)s]+m^2_{\Lambda_b}s[(1-r)^2-s^2]\bigg\}(|A_2|^2+|B_2|^2) \nonumber\\
& & - 8m^2_{\Lambda_b}\bigg\{4m^2_l(1+r-s)-m^2_{\Lambda_b}[(1-r)^2-s^2]\bigg\} (|D_1|^2+|E_1|^2) \nonumber\\
&&+ 8m^5_{\Lambda_b}sv^2\bigg\{-8m_{\Lambda_b}s\sqrt{r}Re(D^*_2E_2) +4(1-r+s)\sqrt{r}Re(D^*_1D_2+E^*_1E_2)\nonumber\\
&& -4(1-r-s) Re(D^*_1E_2+D^*_2E_1)+m_{\Lambda_b}[(1-r)^2-s^2] (|D_2|^2+|E_2|^2)\bigg\},
\end{eqnarray}

\begin{eqnarray}
{\mathcal M}_1(\omega) &=& -16  m_{\Lambda_b}^4 s  v_l \sqrt{\lambda}
\Big\{ 2 Re(A_1^* D_1)-2Re(B_1^* E_1)\nonumber\\
&+& 2m_{\Lambda_b}
Re(B_1^* D_2-B_2^* D_1+A_2^* E_1-A_1^*E_2)\Big\}\nonumber\\
&+&32 m_{\Lambda_b}^5 s  v_l \sqrt{\lambda} \Big\{
m_{\Lambda_b} (1-r)Re(A_2^* D_2 -B_2^* E_2)\nonumber\\
&+&
\sqrt{r} Re(A_2^* D_1+A_1^* D_2-B_2^*E_1-B_1^* E_2)\Big\},
\end{eqnarray}

\begin{eqnarray}
  \mathcal{M}_2(\omega) &=& 8m^6_{\Lambda_b}s v_l^2\lambda(|A_2|^2+|B_2|^2+|E_2|^2+|D_2|^2) \nonumber\\ &- &8 m^4_{\Lambda_b}v_l^2\lambda(|A_1|^2+|B_1|^2+|E_1|^2+|D_1|^2).
\end{eqnarray}

The lepton-side forward-backward asymmetry, $A_{FB}$, is defined as

\begin{eqnarray}
  A_{FB} = \frac{\int_{0}^{1} \frac{d \Gamma}{d q^2 dz} dz -\int_{-1}^{0} \frac{d \Gamma}{d q^2 dz} dz }{\int_{-1}^{1} \frac{d \Gamma}{d q^2 dz} dz },
\end{eqnarray}
where $z = \cos \theta$.
The "naively integrated" observables are obtained by \cite{PRD103-013007}

\begin{eqnarray}
  \langle {X}\rangle &=& \frac{1}{q^2_{max}- q^2_{min}} \int_{q^2_{min}}^{q^2_{max}}X(q^2)d q^2.
\end{eqnarray}

We define the integrated $A_{FB}$ to be
\begin{eqnarray}
  \bar{A}_{FB} &=& \int_{\hat{q}_{min}}^{\hat{q}_{max}} d \hat{q}^2 A_{FB}(\hat{q}^2).
\end{eqnarray}
where $\hat{q}^2= q^2 / M_{\Lambda_b}^2$.
With the aid of the helicity amplitudes of $\Lambda_b \rightarrow \Lambda l^+ l^-$, one can also calculate the hadron forward-backward asymmetry, the lepton-hadron side asymmetry and the fraction of longitudinally polarized dileptons.

The hadron forward-backward asymmetry has the form
\begin{equation}
  \label{EAH}
  A_{FB}^h(q^2)
=\frac{\alpha_\Lambda}{2} \frac{ \frac{v^2_l}{2} ({\cal
    H}_P^{11}+{\cal H}_P^{22}+{\cal H}_{L_P}^{11}+{\cal H}_{L_P}^{22})+\frac{3m_l^2}{q^2}({\cal
    H}_{P}^{11}+{\cal H}_{L_P}^{11}+{\cal H}_{S_P}^{22})}{{\cal H}_{tot}}.\qquad
\end{equation}

The lepton-hadron side asymmetry has the form
\begin{equation}
  \label{ALH}
  A_{FB}^{lh}(q^2)
=-\frac{3}{4} \frac{\alpha_\Lambda}{2} \frac{v_l {\cal H}_U^{12} } {{\cal H}_{tot}}.\qquad
\end{equation}

The fraction of the longitudinally polarized dileptons is expressed by
\begin{equation}
\label{EFL}
F_L(q^2)=\frac{\frac{v^2_l}{2}({\cal
    H}_L^{11}+{\cal H}_{L}^{22})+ \frac{m_l^2}{q^2}({\cal
    H}_{U}^{11}+{\cal H}_{L}^{11}+{\cal H}_{S}^{22})}{{\cal
    H}_{tot}}.
\end{equation}
In Eqs. (\ref{EAH}-\ref{EFL}) ${\cal H}_{X}^{m m^\prime} (X= U,~L,~S,~P,~L_P,~S_P,~m=1,2)$ represent different helicity amplitudes, and ${\cal H}_{tot}$ is the total helicity amplitudes, $\alpha_\Lambda=0.642\pm0.013$. Explicit expression for ${\cal H}^{mm^\prime}_{X}$ can be found in Ref. \cite{PRD96-053006}.

\section{Numerical analysis and discussion}

In this section we perform a detailed numerical analysis of $A_{FB}( \Lambda_b \rightarrow \Lambda l^+ l^-)$.
In this work, the masses of baryons, $m_{\Lambda_b}=5.62$ GeV and $m_\Lambda=1.116$ GeV \cite{PDG2020}, the masses of quarks, $m_b=5.02$ GeV and $m_s=0.516$ GeV \cite{PRD95-054001, PRD87-076013, PRD91-016006}, are taken.
The variable $\omega$ changes from $1$ to $2.617,~2.614,~1.617$ for $e,~\mu,~\tau$, respectively.

Solving Eqs. (\ref{BS:f12}) and (\ref{BS:hq}) for $\Lambda$ and $\Lambda_b$ we can get the numerical solutions of their BS wave functions.
In Table. \ref{TB:alpha}, we give the values of $\alpha_{seff}$ with different values of $\kappa $ for $\Lambda$ and $\Lambda_b$ with $E_0=-0.14$ GeV.
%while in Fig. \ref{Fig:B} - \ref{Fig:k}, we plot the BS wave functions for $\Lambda_b$ and $\Lambda$.

\begin{table}[!htb]
\centering  % 表居中
\begin{tabular}{c||c|c|c|c|c|c}  % {lccc} 表示各列元素对齐方式，left-l,right-r,center-c
\hline
  $\kappa$ (GeV$^3$)  & 0.045  & 0.047 &0.049 &0.051 &0.053  &0.055 \\ \hline \hline% 在此行下面画一横线
$\Lambda$ &0.559 &0.555 & 0.551& 0.547& 0.544& 0.540 \\  \hline       % \\ 表示重新开始一行
$\Lambda_b$  &0.775 & 0.777&0.778 &0.780& 0.782 &0.784 \\
 \hline
\end{tabular}
\caption{The values of $\alpha_{seff}$  for $\Lambda$ and $\Lambda_b$ with different $\kappa$.}\label{TB:alpha}
\end{table}

%
%\begin{figure}[tb]
%\begin{center}
%\begin{tabular}{ccc}
%%\vspace{-2cm}
%\includegraphics[width=6cm]{Mf14kappa.eps}
%\includegraphics[width=6cm]{Mf005e0.eps}
%\put (-320,-70){(a)~$E_0=-0.14$GeV} \put (-120,-70){(b)~$\kappa=0.05$GeV$^3$}
%\end{tabular}
%%\vskip 2cm
%\caption{(color online ) The BS wave functions for $\Lambda_b$.}\label{Fig:B}
%\end{center}
%\end{figure}
%
%
%
%
%\begin{figure}[tb]
%\begin{center}
%\begin{tabular}{ccc}
%%\vspace{-2cm}
%\includegraphics[width=6cm]{f114kp.eps}
%\includegraphics[width=6cm]{f214kp.eps}
%\end{tabular}
%\vskip 2cm
%\caption{(color online ) The BS wave functions for $\Lambda_b$ with $E_0=-0.14$GeV.}\label{Fig:s}
%\end{center}
%\end{figure}
%
%
%
%\begin{figure}[tb]
%\begin{center}
%\begin{tabular}{ccc}
%%\vspace{-2cm}
%\includegraphics[width=6cm]{f1005e0.eps}
%\includegraphics[width=6cm]{f2005e0.eps}
%\end{tabular}
%\vskip 2cm
%\caption{(color online ) The BS wave functions for $\Lambda_b$ with $\kappa=-0.05$GeV$^3$.}\label{Fig:k}
%\end{center}
%\end{figure}

From Table \ref{TB:alpha}, we find that the value of $\alpha_{seff}$ is weakly dependent on the value of $\kappa$.
%From the figures in Figs. \ref{Fig:B} - \ref{Fig:k}, we find that the BS wave functions of $\Lambda$ are very similar for different $\kappa$, and the values of $\phi(p_t)$, $f_1(p_t)$ and $f_2(p_t)$  change from $0$ to $0.16$, change from $0$ to $0.15$ and $f_2(p_t)$ $0$ to $0.02$, respectively.
%In Fig. \ref{FFs-fig}, we give the FFs and $R(\omega)=F_2/F_1$ for different values of $\kappa$.
From this figure, we find that $R(\omega)$ varies from $-0.75$ to $-0.25$ in our model.
In Ref. \cite{PRD59-114022} $R(\omega)$ varies from $-0.42$ to $-0.83$ in the same $\omega$ region which is in agreement with our result and the estimated value from Refs. \cite{CPC44-083107, EPJC80-193} mentioned in Introduction.
In the range of $2.43 \leq \omega \leq 2.52$ (corresponding to $M_\Lambda^2 \leq q^2 \leq M_{\Lambda_c}^2$), $R(\omega)$ is about $-0.25$.
In the same $\omega$ region, assuming the FFs have the same dependence on $q^2$, the CLEO Collaboration measured $R=-0.35\pm 0.04\pm0.04$ in the limit $m_c \rightarrow + \infty$.
These results are in good agreement our work in the same $\omega$ region.

\begin{table}[!htb]
\centering  % 表居中
\begin{tabular}{ c|c|c|c|c }  % {lccc} 表示各列元素对齐方式，left-l,right-r,center-c
\hline
    & $\bar{A}^l_{FB}$ & $\bar{A}^{lh}_{FB}$  & $\bar{A}^{h}_{FB}$ & $\bar{F}_L$   \ \\  \hline
 \cite{PRD64-074001, PLB516-327} &$-0.13$& - &-  &$0.5830$  \ \\
\cite{PRD87-074031} &$8.0\times 10^{-4}$& - &-  &-     \ \\
\cite{PRD96-053006} &  $-0.286$& $0.101$ & $-0.288$ &$0.525$ \ \\
\cite{EPJC59-861} & $-0.0122^{+0.0142}_{-0.0073}$& - & -&- \ \\
\cite{JHEP01-155} &$-0.29\pm0.05$& $0.13^{+0.22}_{-0.03}$  & $-0.26\pm0.03$ &$0.4\pm0.1$  \ \\
\cite{PRD103-013007} &$-0.04^{+0.00}_{-0.01}$& - &-  &$0.34_{-0.02}^{+0.03}$    \ \\
our work  &$-0.1376\pm0.0001$& $0.0576$ &$- 0.1613\pm0.0001$ &$0.3957\pm0.0002$   \ \\ \hline
\end{tabular}
\caption{ Longitudinal polarization fractions and forward-backward asymmetries for $\Lambda_b \rightarrow \Lambda \mu^+ \mu^-$.}\label{tabu}
\end{table}

\begin{table}[!htb]
\centering  % 表居中
\begin{tabular}{ c|c|c|c|c  }  % {lccc} 表示各列元素对齐方式，left-l,right-r,center-c
\hline
 -& $A^l_{FB [15,20]}$   & $ {A}^{lh}_{FB[15,20]}$  & $ {A}^{h}_{FB[15,20]}$ & $ {F}_{L[15,20]}$  \ \\  \hline
LHCb \cite{JHEP09-146,JHEP06-115}& $-0.39\pm0.04$  & - &$-0.29\pm0.07 $& $0.61^{+0.11}_{-0.14}$    \ \\
 \cite{PRD64-074001, PLB516-327}&$-0.40\sim-0.25$ &-& - &$0.37\sim 0.62$    \ \\
\cite{PRD87-074031}&$-0.24\sim -0.13$&-& $>-0.308$ &-  \ \\
\cite{PRD96-053006}&$-0.40$&$0.145$ &  $-0.29$ & $0.38$    \ \\
\cite{EPJC59-861} & $-0.075\sim -0.017$&-&-&- \ \\
\cite{PRD103-013007}&$-0.34_{-0.02}^{+0.01}$ &-& - &  $0.4^{+0.01}_{-0.02}$   \ \\
\cite{PRD93-074501}&$-0.350(13)$&-& $-0.2710\pm0.0092$ &$0.409\pm0.013$     \ \\
our work &$-0.44\sim-0.35$&$0.1257\sim 0.1555$ & $-0.2304\sim-0.0685$  & $0.3398\sim0.4530$   \ \\ \hline
\end{tabular}
\caption{ Longitudinal polarization fractions and forward-backward asymmetries for $\Lambda_b \rightarrow \Lambda \mu^+ \mu^-$ in $q^2 \in [15,20]$ GeV$^2$.}\label{tabubin}
\end{table}

In Table \ref{tabu}, we give $ \bar{A}^l_{BF}$, $\bar{A}^{lh}_{FB}$, $\bar{A}^h_{FB}$ and $\bar{F}_L$ for $\Lambda_b \rightarrow \Lambda \mu^+ \mu^-$ and compare our results with other works.
We can see that these asymmetries differ a lot in different models.
Considering these differences, $\bar{A}^l_{FB}$ changes between $-0.30$ and $0$, $\bar{A}^{lh}_{FB}$ is about $0.1$, $\bar{A}^h_{FB}$ is about $-0.25$, and $\bar{F}_L$ changes from $0.3$ to $ 0.6$.
Without including the long distance contribution, Ref. \cite{PRD64-074001} gave the integrated forward-backward asymmetry $\bar{A}^l_{BF}(\Lambda_b \rightarrow \Lambda \mu^+ \mu^-)=-0.1338$.
The result of Ref. [10] were $\bar{A}^l_{BF}(\Lambda_b \rightarrow \Lambda \mu^+ \mu^-)=-0.13(-0.12)$ in the QCD sum rule approach (the pole model).
Using the covariant constituent quark model with (without) the long distance contribution, Ref. \cite{PRD87-074031} gave the result $\bar{A}^l_{BF}(\Lambda_b \rightarrow \Lambda \mu^+ \mu^-)=1.7\times 10^{-4} (8\times 10^{-4})$.

\begin{figure}[htbp]
\begin{center}
\begin{minipage}[t]{0.3\linewidth}
 \includegraphics[width=5.0cm]{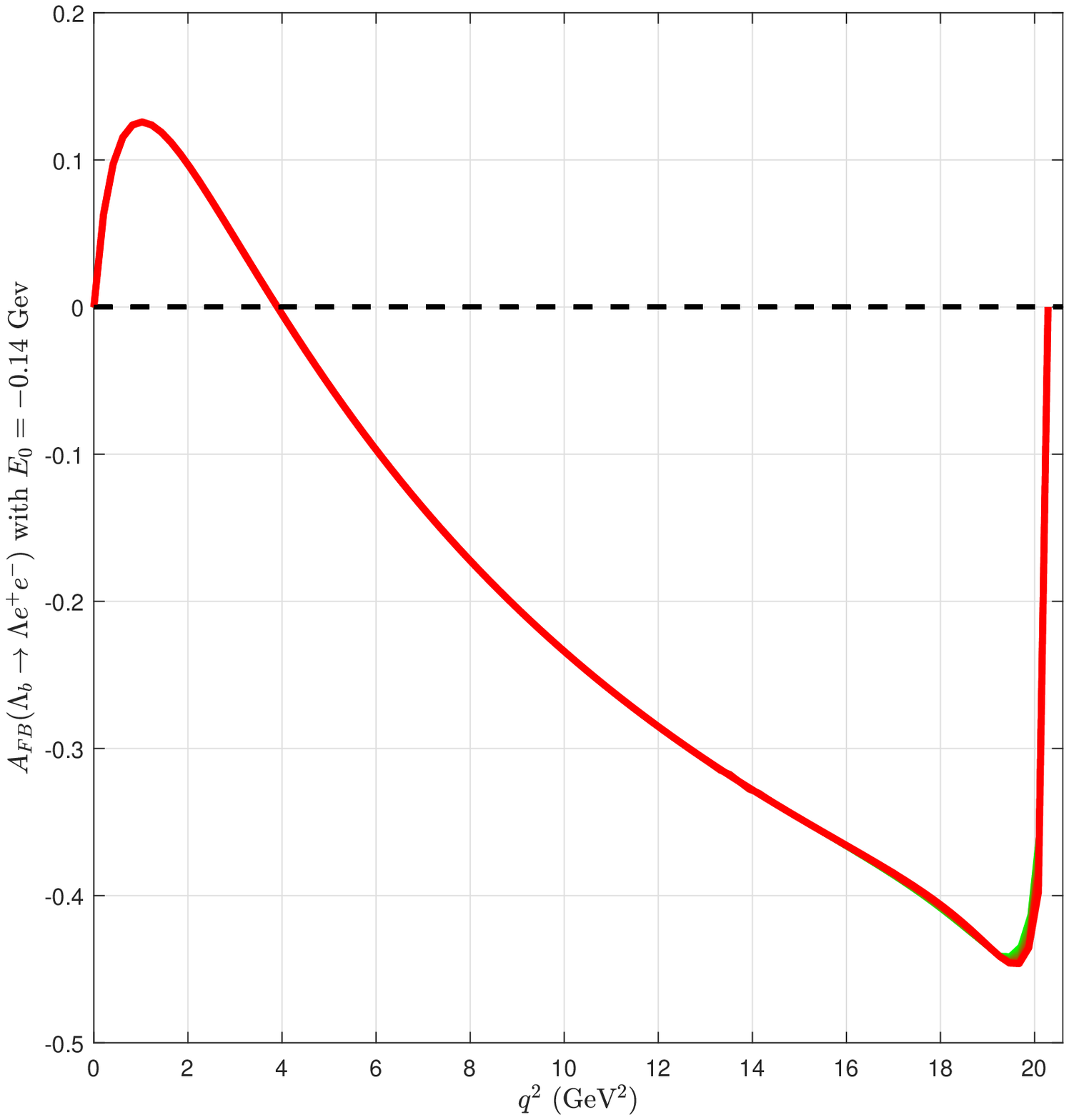}
\end{minipage}
 \begin{minipage}[t]{0.3\linewidth}
 \includegraphics[width=5.0cm]{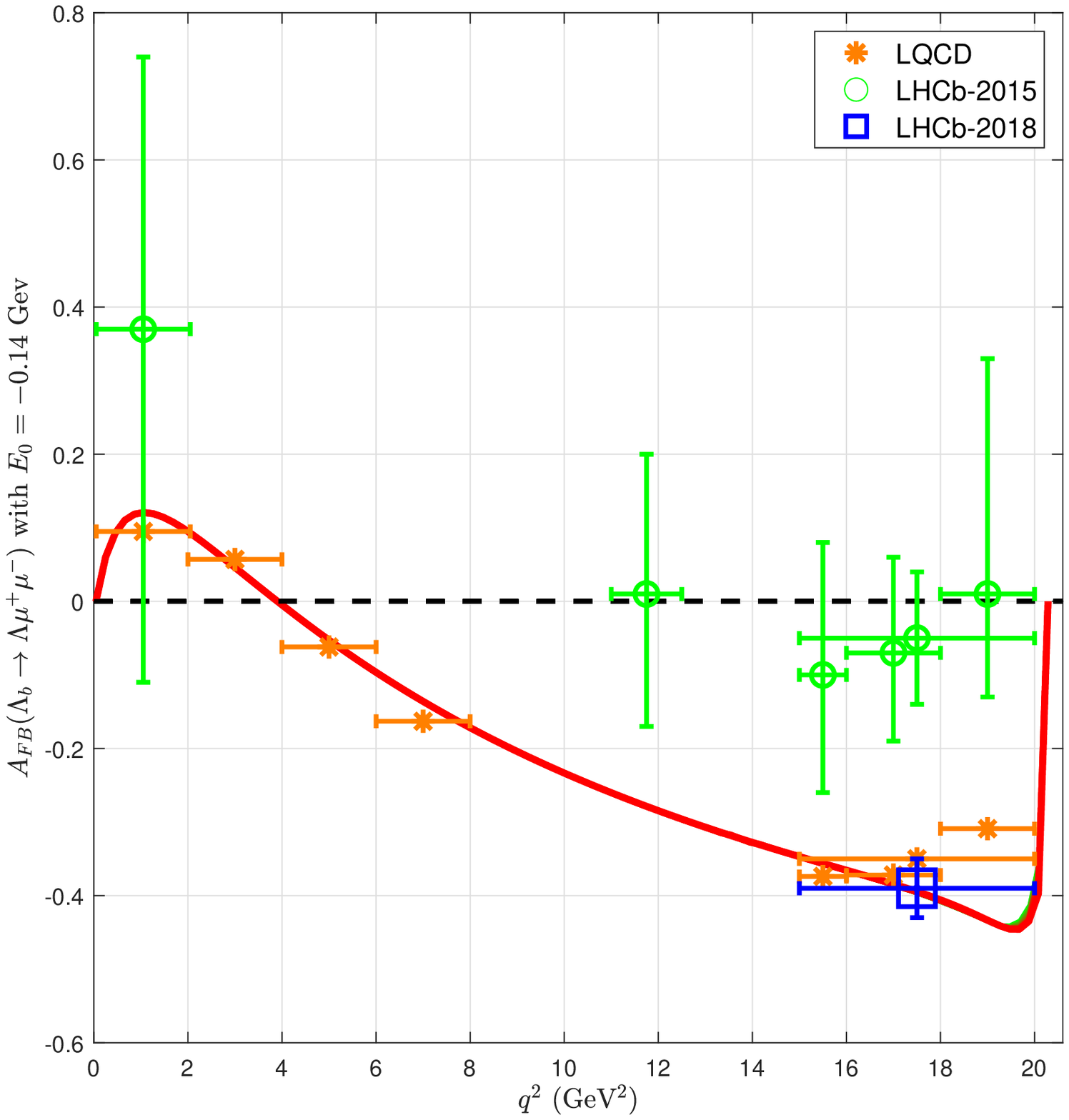}
\end{minipage}
\begin{minipage}[t]{0.3\linewidth}
 \includegraphics[width=5.0cm]{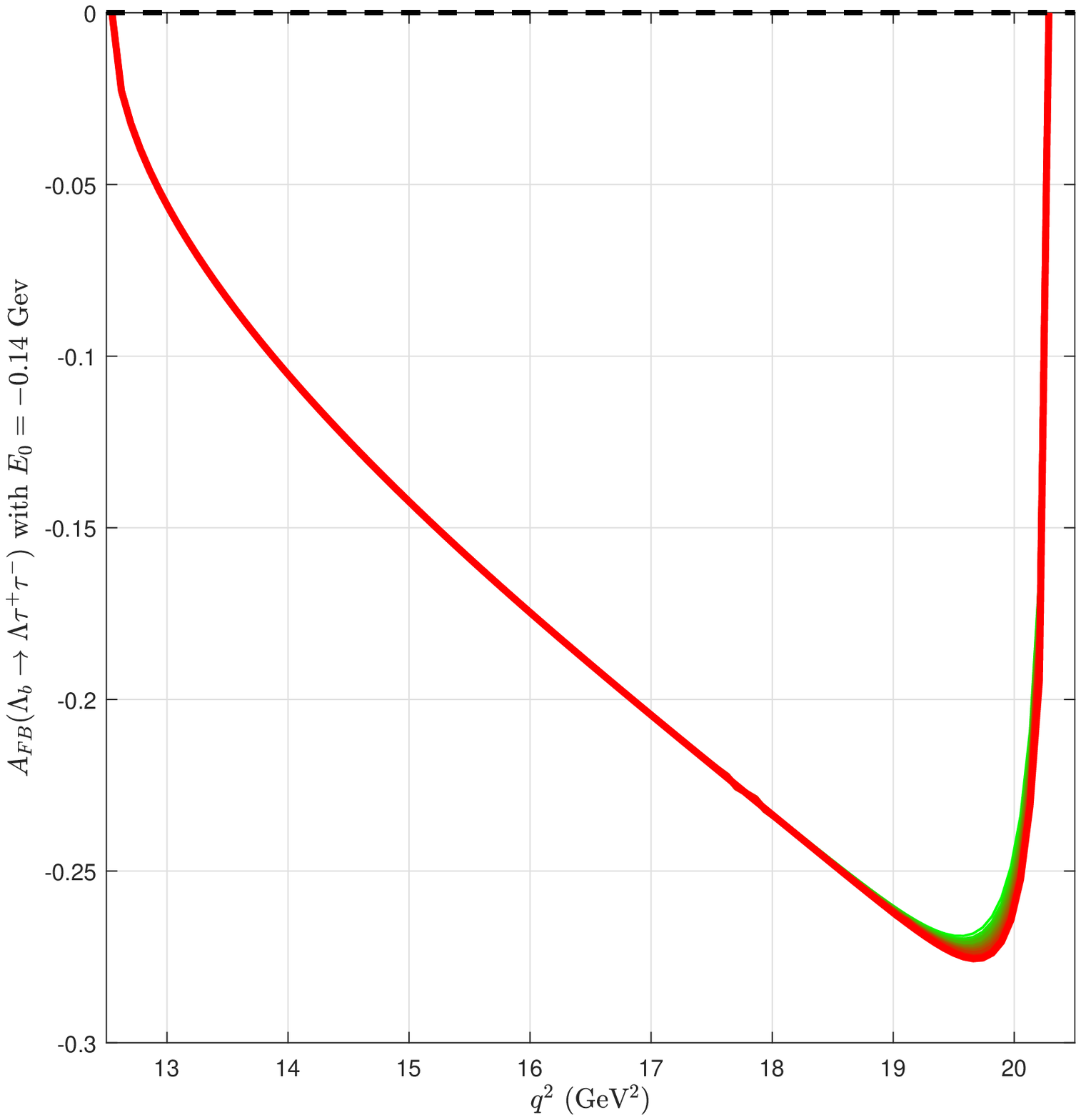}
\end{minipage}
\caption{(color online) Values of $A_{FB}(\Lambda_b\rightarrow\Lambda l^+ l^-)$ as a functions of $q^2$ for different values of $\kappa$ as shown in Table \ref{TB:alpha}.}\label{FBA}
\end{center}
\end{figure}

\begin{figure*}[!htbp]
\begin{center}
\begin{minipage}[t]{0.3\linewidth}
 \includegraphics[width=5.0cm]{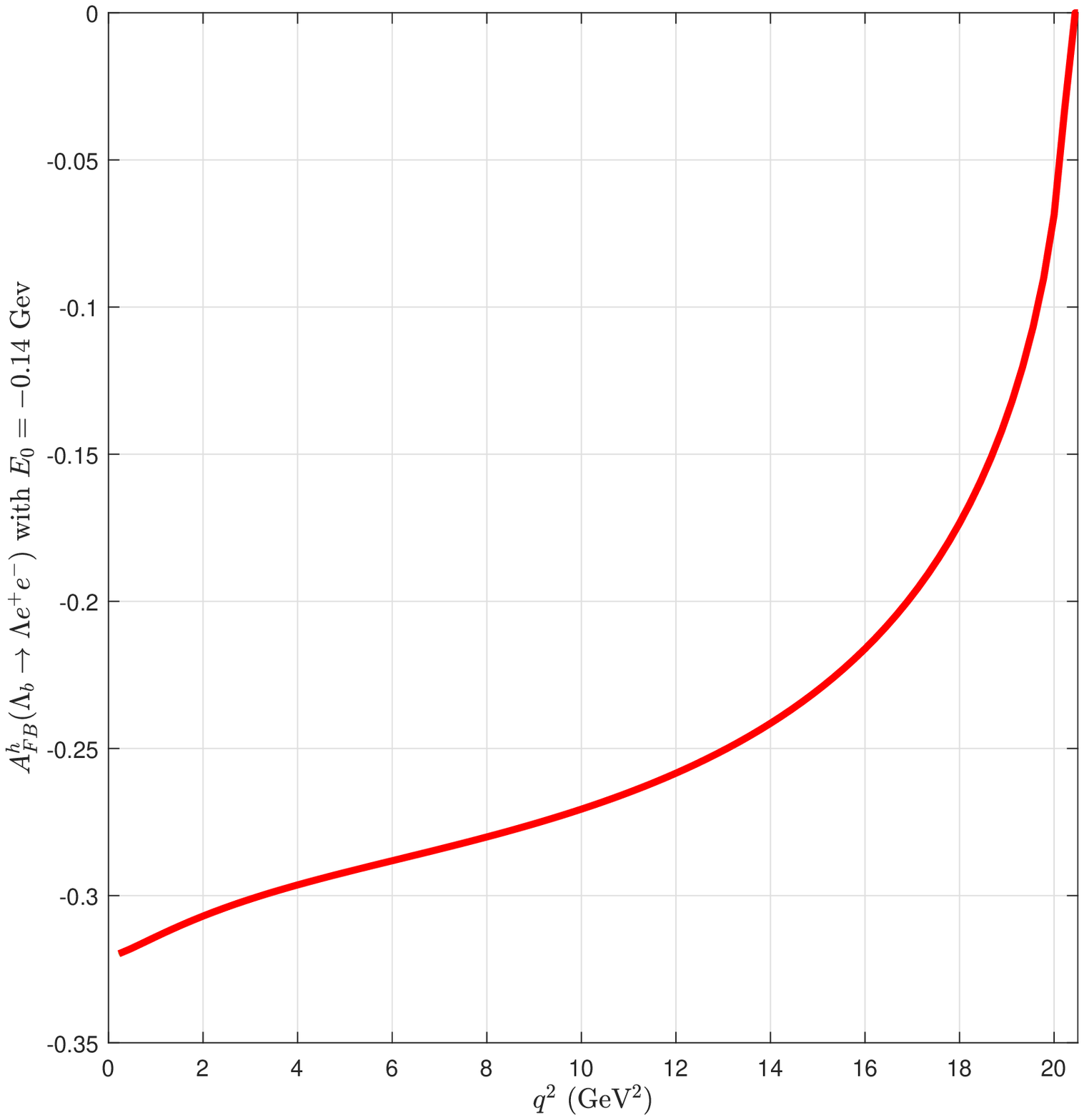}
\end{minipage}
\begin{minipage}[t]{0.3\linewidth}
 \includegraphics[width=5.0cm]{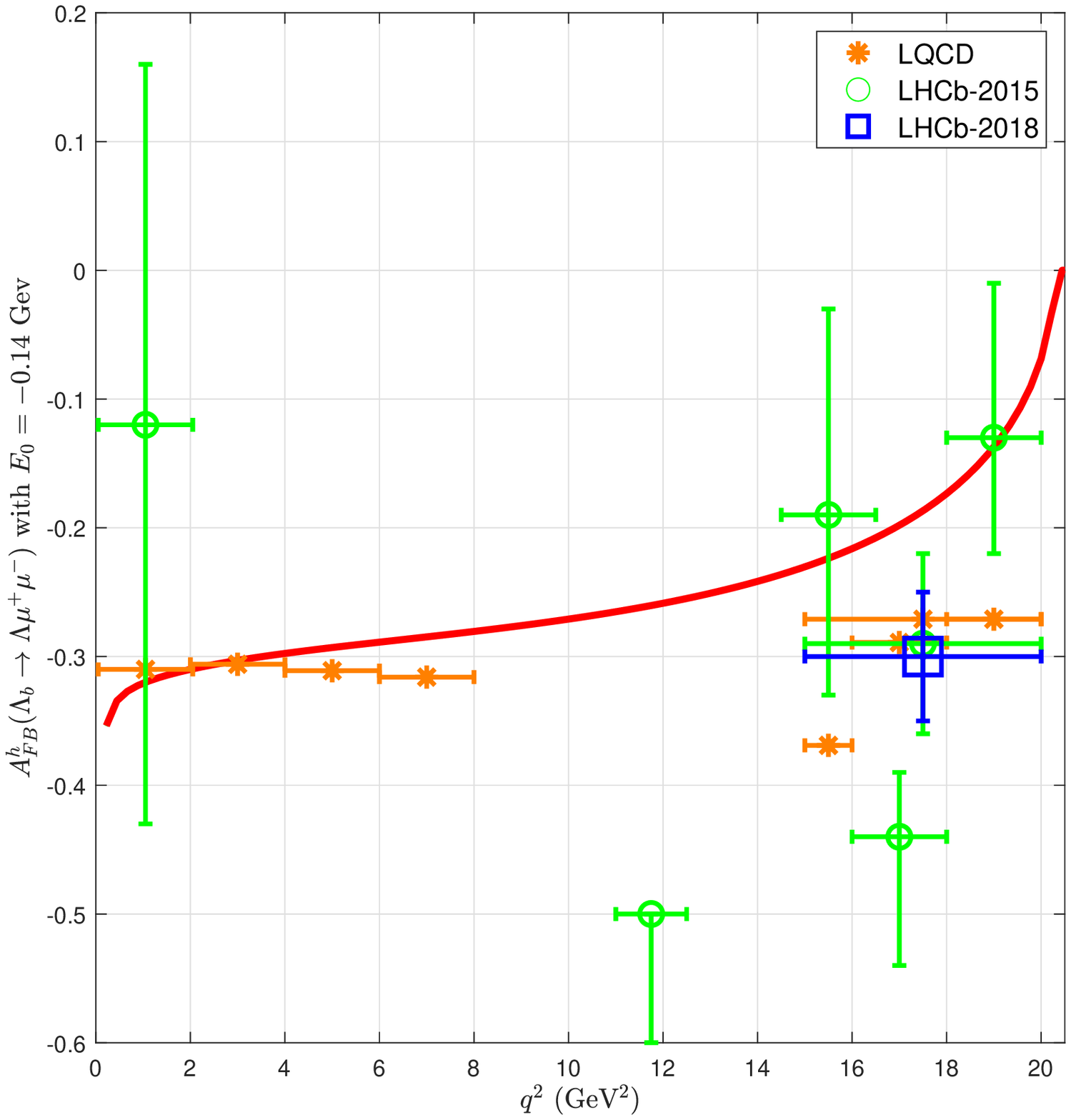}
\end{minipage}
\begin{minipage}[t]{0.3\linewidth}
 \includegraphics[width=5.0cm]{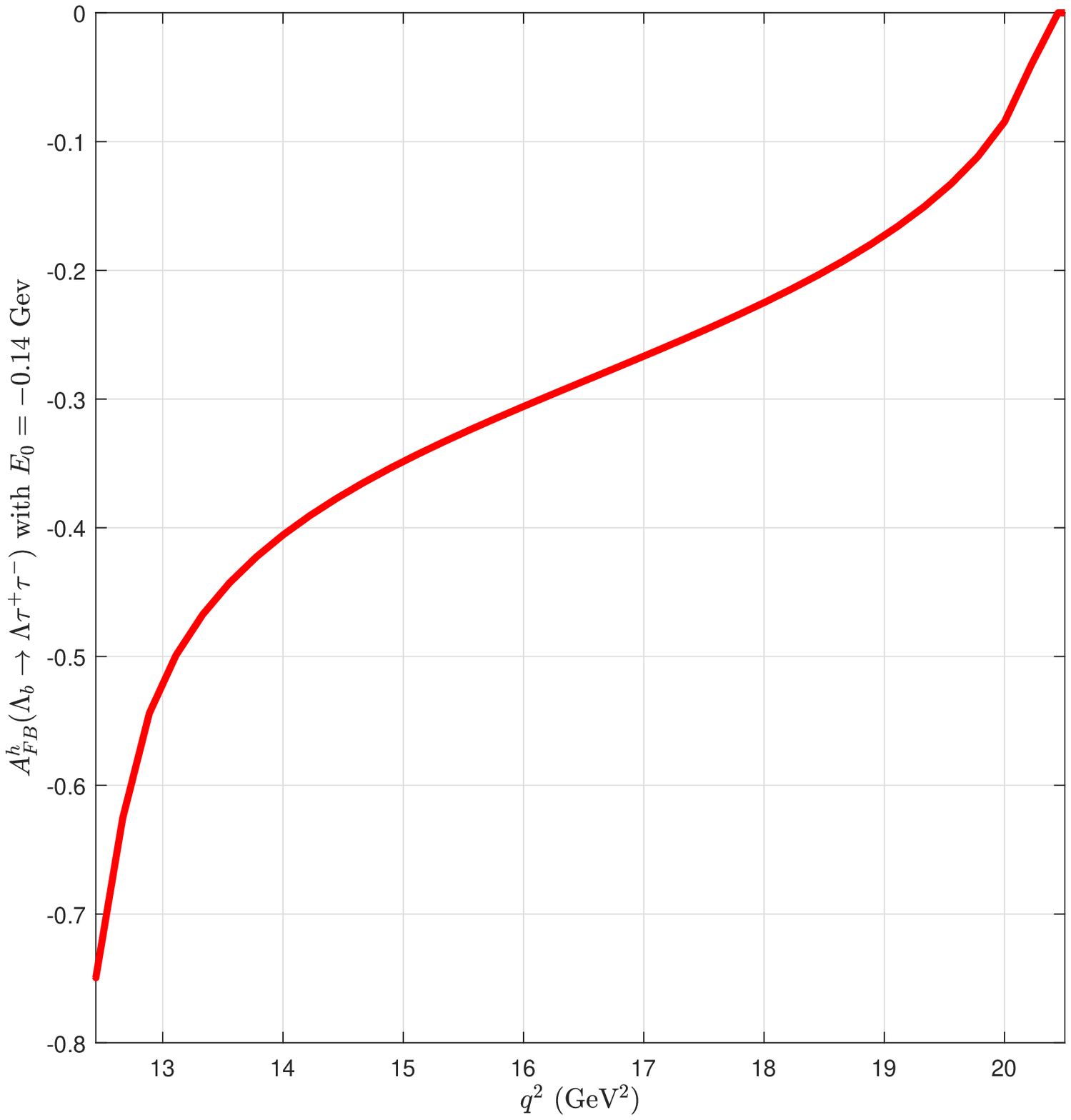}
\end{minipage}
\caption{(color online) Values of $A^h_{FB}(\Lambda_b\rightarrow\Lambda l^+ l^-)$ as a functions of $q^2$ for different values of $\kappa$ as shown in Table \ref{TB:alpha}.}\label{HS}
\end{center}
\end{figure*}

\begin{figure*}[!htbp]
\begin{center}
\begin{minipage}[t]{0.3\linewidth}
 \includegraphics[width=5.0cm]{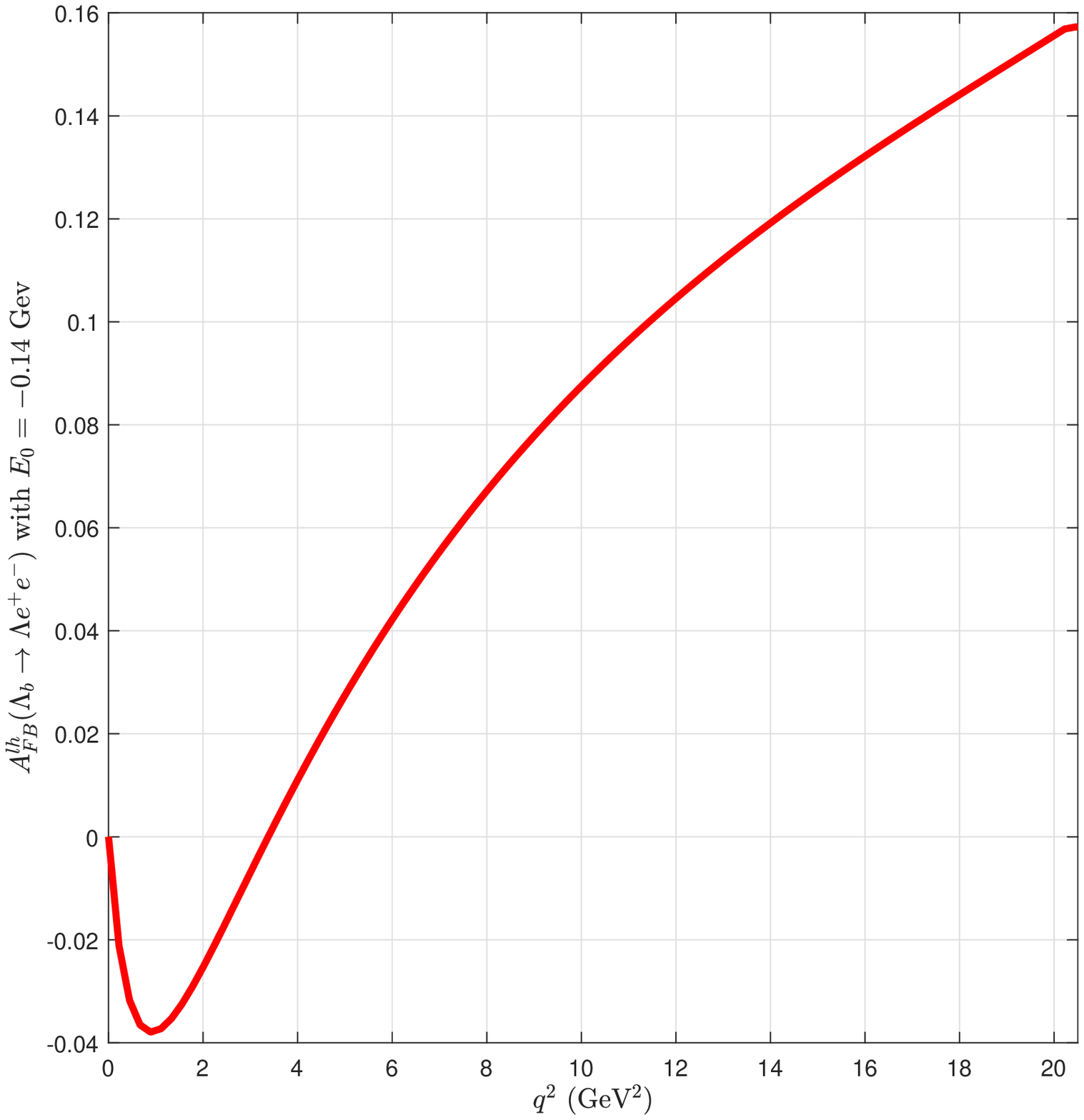}
\end{minipage}
\begin{minipage}[t]{0.3\linewidth}
 \includegraphics[width=5.0cm]{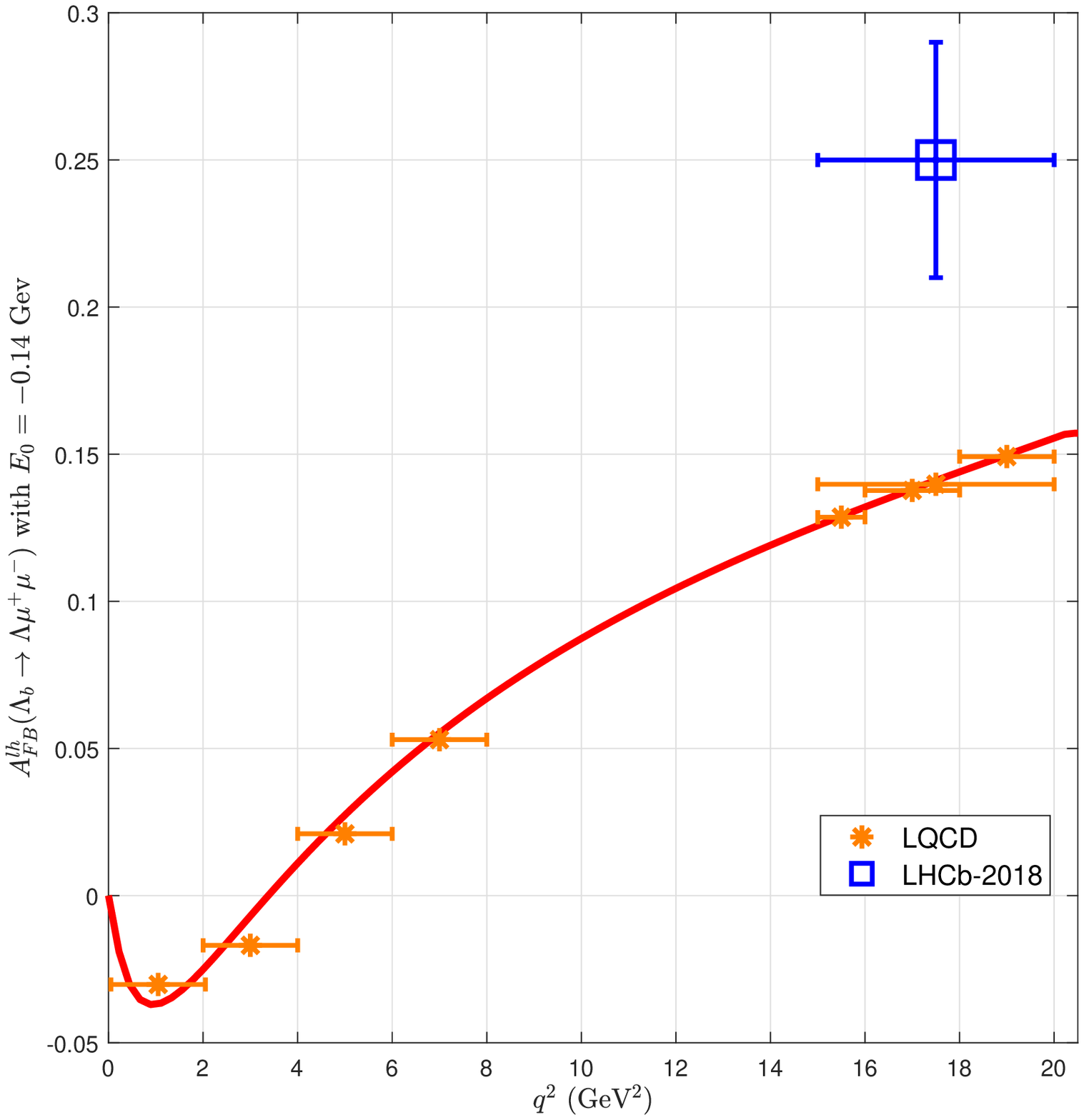}
\end{minipage}
\begin{minipage}[t]{0.3\linewidth}
 \includegraphics[width=5.0cm]{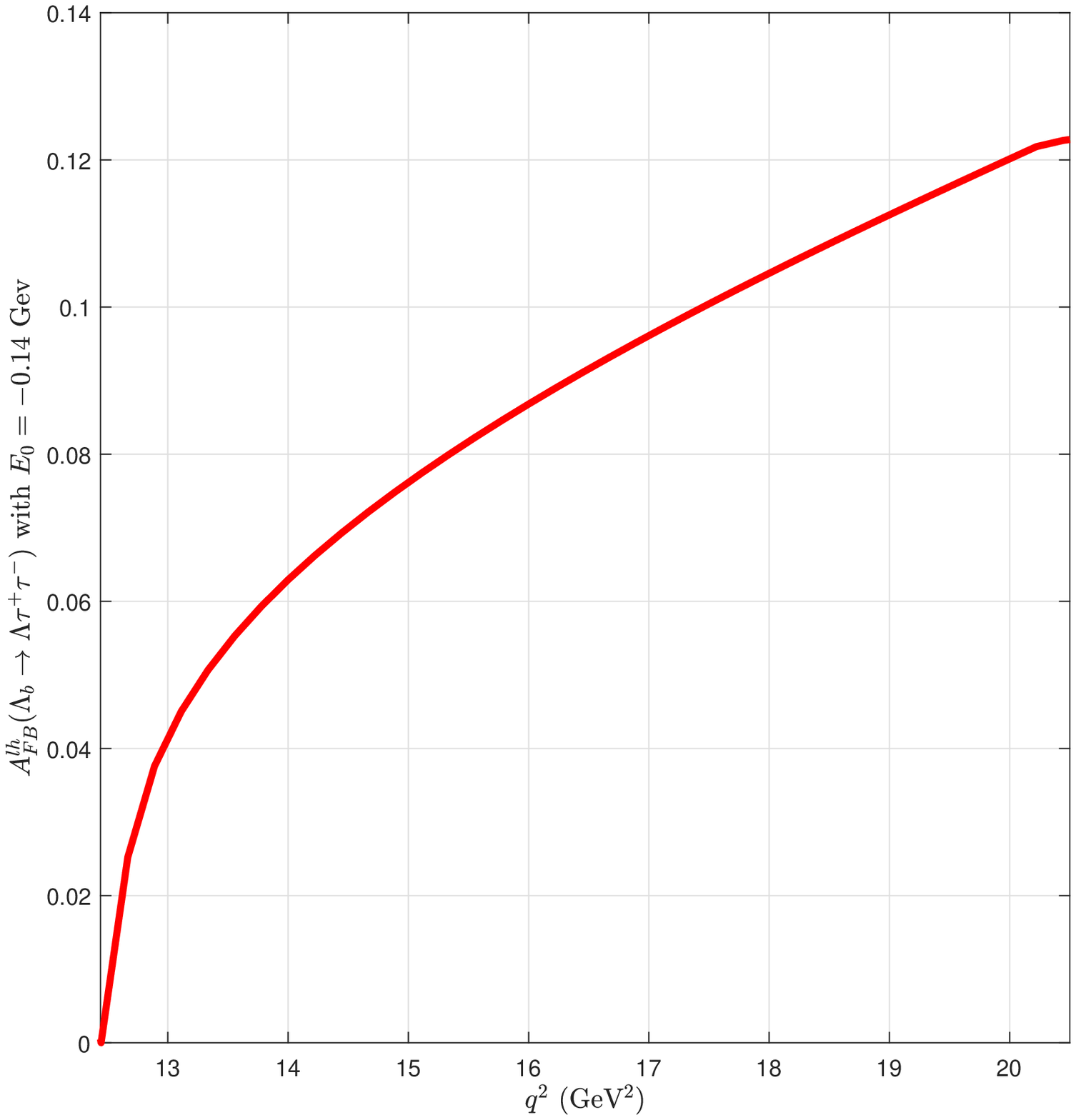}
\end{minipage}
\caption{(color online) Values of $A^h_{FB}(\Lambda_b\rightarrow\Lambda l^+ l^-)$ as a functions of $q^2$ for different values of $\kappa$ as shown in Table \ref{TB:alpha}.}\label{LH}
\end{center}
\end{figure*}

\begin{figure*}[!htbp]
\begin{center}
\begin{minipage}[t]{0.3\linewidth}
 \includegraphics[width=5.0cm]{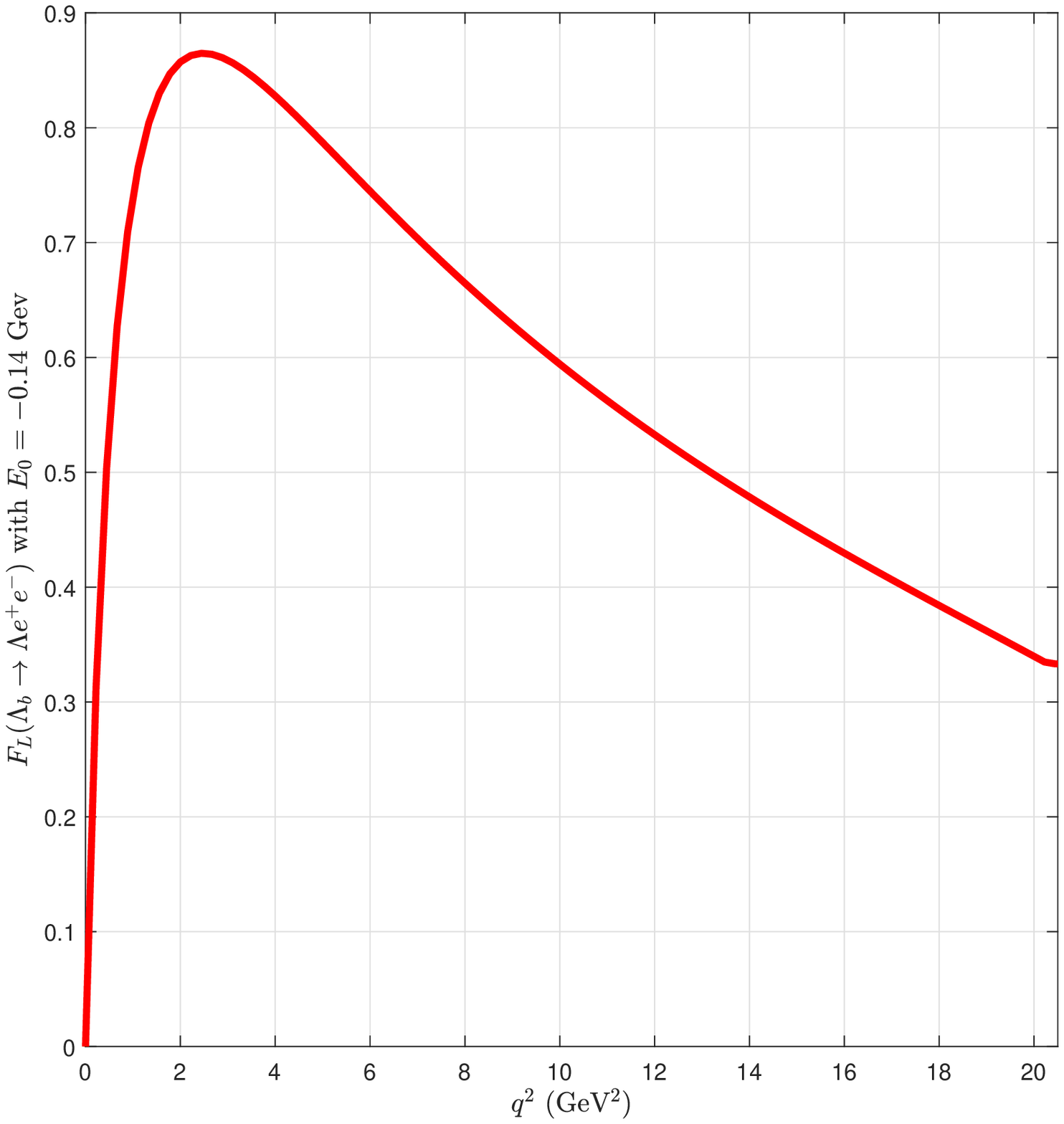}
\end{minipage}
\begin{minipage}[t]{0.3\linewidth}
 \includegraphics[width=5.0cm]{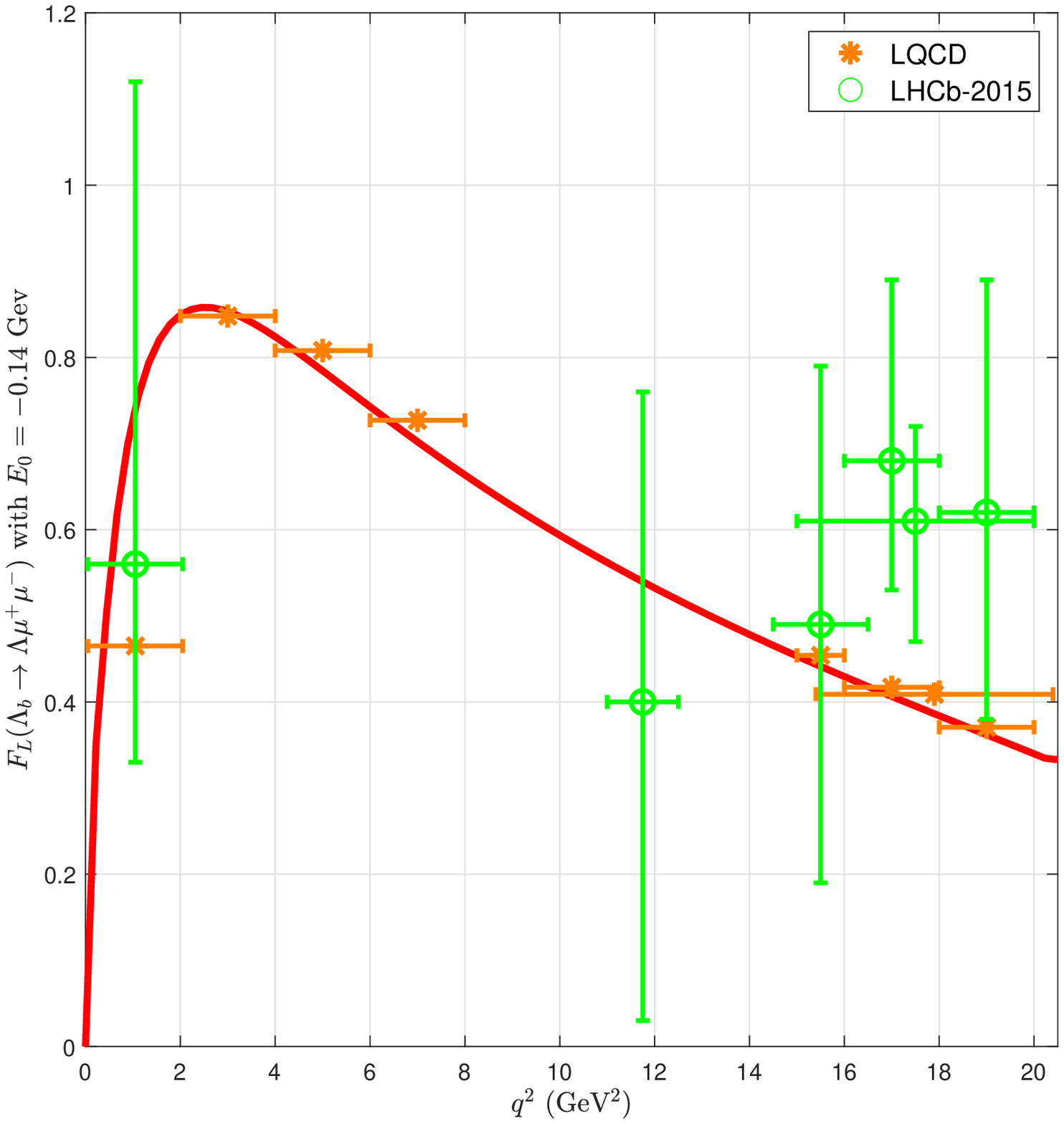}
\end{minipage}
\begin{minipage}[t]{0.3\linewidth}
 \includegraphics[width=5.0cm]{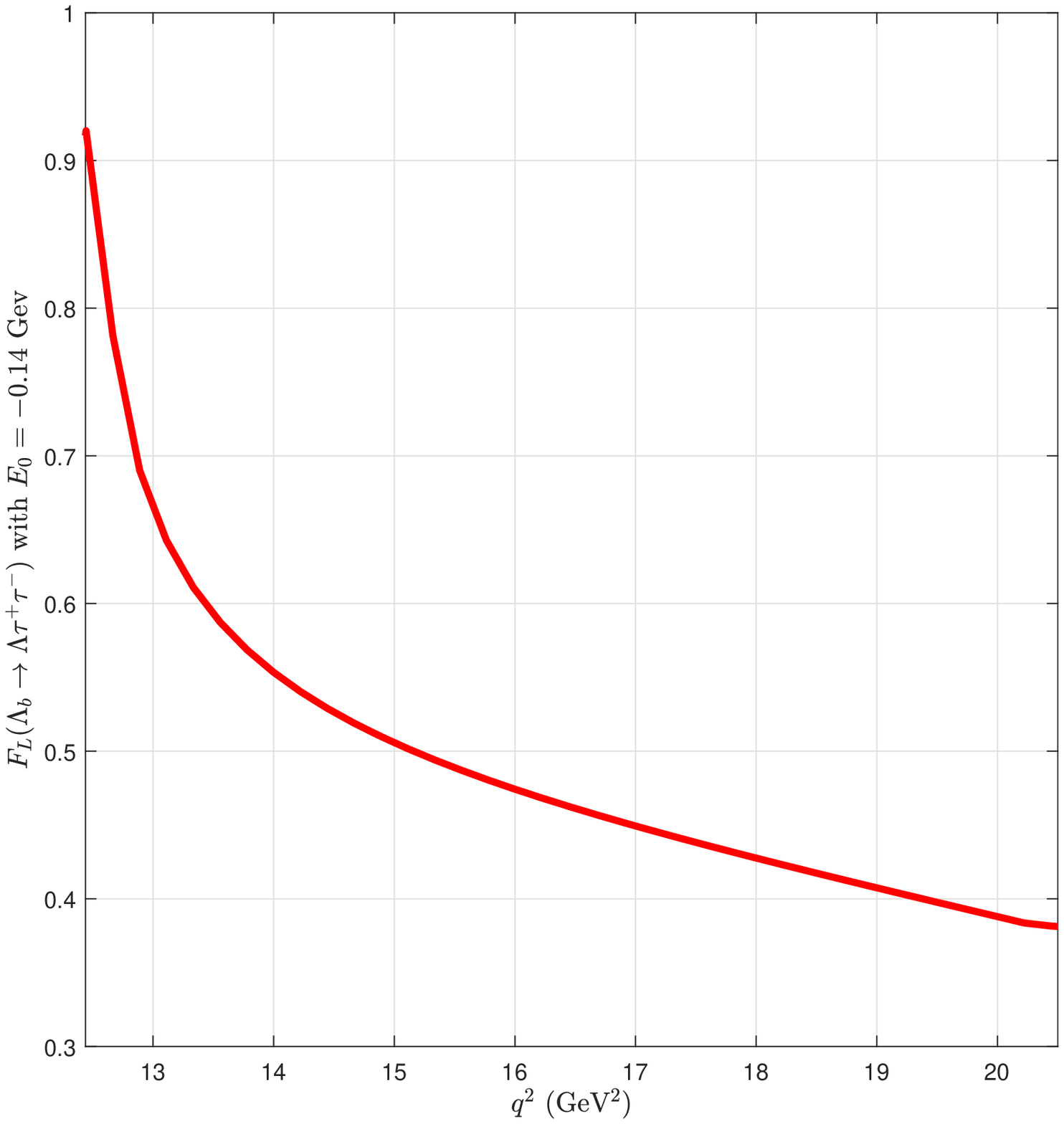}
\end{minipage}
\caption{(color online) Values of $F_L(\Lambda_b\rightarrow\Lambda l^+ l^-)$ as a functions of $q^2$ for different values of $\kappa$ as shown in Table \ref{TB:alpha}.}\label{FL}
\end{center}
\end{figure*}

When $q^2 \in [15,20]$ GeV$^2$, the LHCb Collaboration gave ${A}^l_{FB}(\Lambda_b \rightarrow \Lambda \mu^- \mu^+) = -0.05 \pm 0.09$ in 2015 which was updated to be ${A}^l_{FB}(\Lambda_b \rightarrow \Lambda \mu^- \mu^+) = -0.39 \pm 0.04$ three years later \cite{JHEP09-146, JHEP06-115}.
In our work, in the same region the value of $ {A}^l_{BF}(\Lambda_b \rightarrow \Lambda \mu^- \mu^+)$ changes from $-0.44$ to $-0.35$ which is in good agreement with the most recent experimental data of LHCb.
With the latest high-precision lattice QCD calculations in the same region, Ref. \cite{Arxiv:1701-04029} gave the values ${A}^l_{FB}(\Lambda_b \rightarrow \Lambda \mu^- \mu^+) = -0.344$ in the large $\varsigma_u$ and small $\varsigma_d$ region ($\varsigma_u,~ \varsigma_d$ are model parameters \cite{EPJC77-190}) and $ {A}^l_{FB}(\Lambda_b \rightarrow \Lambda \mu^- \mu^+) =-0.24$ in the large $\varsigma_d$ and small $\varsigma_u$ region.
In Figs. \ref{FBA}, we plot the $q^2$-dependence of $A^l_{FB}(\Lambda_b \rightarrow \Lambda e^- e^+) $, $A^l_{FB}(\Lambda_b \rightarrow \Lambda \mu^- \mu^+) $ and $A^l_{FB}(\Lambda_b \rightarrow \Lambda \tau^- \tau^+) $.
From Fig. \ref{FBA}, we can see that $A^l_{FB}(\Lambda_b\rightarrow\Lambda \mu^+ \mu^-)$ is in good agreement with LQCD in all the $q^2$ region \cite{PRD93-074501}.
The results of other references results are also shown in Table \ref{tabubin}.
In Figs. \ref{HS}, we plot the $q^2$-dependence of $A^h_{FB}(\Lambda_b \rightarrow \Lambda e^- e^+) $, $A^h_{FB}(\Lambda_b \rightarrow \Lambda \mu^- \mu^+) $ and $A^h_{FB}(\Lambda_b \rightarrow \Lambda \tau^- \tau^+)$, respectively.
When $q^2 \in [15,20]$ GeV$^2$, the LHCb Collaboration gave the value for $\Lambda_b \rightarrow \Lambda \mu^- \mu^+$ as $-0.29\pm0.07$ which is in good agreement our result $-0.2304\sim -0.0685$.
The results of other references results are also shown in Table \ref{tabubin}.
In Figs. \ref{LH},we plot the $q^2$-dependence of $A^{lh}_{FB}(\Lambda_b \rightarrow \Lambda e^- e^+) $, $A^{lh}_{FB}(\Lambda_b \rightarrow \Lambda \mu^- \mu^+) $ and $A^{lh}_{FB}(\Lambda_b \rightarrow \Lambda \tau^- \tau^+)$, respectively.
Ref. \cite{PRD96-053006} gave the value  $A^{lh}_{FB}(\Lambda_b \rightarrow \Lambda \mu^- \mu^+) = 0.145$ which is agreement with our results $0.1257\sim0.1555$ in the region $q^2 \in [15,20]$ GeV$^2$.
In Figs. \ref{FL}, we plot the $q^2$-dependence of $F_L(\Lambda_b \rightarrow \Lambda e^- e^+) $, $F_L(\Lambda_b \rightarrow \Lambda \mu^- \mu^+) $ and $F_L(\Lambda_b \rightarrow \Lambda \tau^- \tau^+) $, respectively.
In the region $q^2 \in [15,20]$ GeV$^2$, the LHCb Collaboration gave the value $F_L(\Lambda_b \rightarrow \Lambda \mu^- \mu^+)=0.61_{-0.14}^{+0.11} $ which is close to our result $0.3398\sim0.4530$.
The results of other references results are also shown in Table \ref{tabubin}.
From these figures, we find that all these asymmetries are not very sensitive to the parameters $\kappa $ and $E_0$ in our model.

\begin{figure}[htbp]
\begin{center}
 \includegraphics[width=8.0cm]{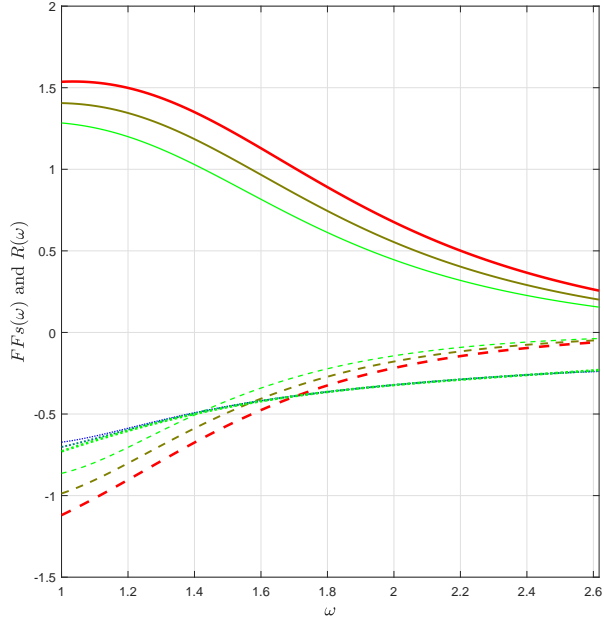}
\caption{(color online) Values of $F_1$ (solid line), $F_2$ (dash line) and $R(\omega)$ (dot line) as a function of $\omega$ (the lines become thicker with the increase of $\kappa$)}\label{FFs-fig}
\end{center}
\end{figure}

Ref. \cite{PRD103-013007} gave the naively integrated values $\langle A_{FB}^l \rangle  = -0.19^{+0.00}_{-0.01}$ and $\langle F_L \rangle = 0.6\pm0.02$ for $\Lambda_b \rightarrow \Lambda \mu^+ \mu^-$, while in our work these values are $-0.1976$ and $0.5681$, respectively.
 Obviously, our results are very close to those of Ref.\cite{PRD103-013007}.
  In our work, we give $\bar{A}^l_{FB}=-0.0708\pm0.0001(-0.0590\pm0.0001)$ and $\bar{A}^{h}_{FB}=-0.1604\pm0.0001(-0.1541\pm0.0002)$ for $\Lambda_b\rightarrow\Lambda e^+ e^-(\Lambda_b\rightarrow\Lambda \tau^+ \tau^-)$.
  The values given in Ref. \cite{PRD87-074031} are $\bar{A}^l_{FB}=1.2\times 10^{-8}(9.6\times 10^{-4})$ and $\bar{A}^{h}_{FB}=-0.321(-0.259)$, while Refs. \cite{EPJC59-861} and Ref. \cite{PLB516-327} gave $\bar{A}^l_{FB}=-0.0067$ and $\bar{A}^l_{FB}=-0.04 $ for $\Lambda_b\rightarrow\Lambda \tau^+ \tau^-$.
  Comparing  the values in these theoretical approaches we find that the asymmetries may vary widely among the theoretical models because the FFs in these models are different.

\section{summary and conclusions}

  In the present work, we used the BSE to study the forward-backward asymmetries in the rare decays $\Lambda_b \rightarrow \Lambda l^+ l^-$ in a covariant quark-diquark model.
  In this picture, $\Lambda_b (\Lambda) $ is regarded as a bound state of a $b(s)$-quark and a scalar diquark.

  We established the BSE for the quark and the scalar diquark system and then we derived the FFs of $\Lambda_b \rightarrow \Lambda $.
  We solved the BS equation of this system and then gave the values of the FFs and $R$.
  We found that the ratio $R$ is not a constant which is in agreement with Ref. \cite{PRD53-4946}  and the pQCD scaling law \cite{PRD22-2157, PPNP59-694, PRD11-1309}.
  Using these FFs, we calculated the forward-backward asymmetries $A^l_{FB}$, $A^{lh}_{FB}$, $A^h_{FB}$ and the longitudinal polarization fractions $F_L$ and the integrated forward-backward asymmetries $\bar{A}^l_{FB}$, $\bar{A}^{lh}_{FB}$, $\bar{A}^h_{FB}$ and $\bar{F}_L$ for $\Lambda_b \rightarrow \Lambda l^+l^- (l=e,~\mu,~\tau)$.
  Comparing with other theoretical works we found that the FFs are different, thus these asymmetries are different.
The long distance contributions is not included in our present work, in order to compare with the experimental data more exactly will be considered in our future work.

%%%%%%%%%%%%%%%%%%%%%%%%%%%%%%%%%%%%%%%%%%%%%%%%%%%%%%%%%
\acknowledgments
This work was supported by National Natural Science Foundation of China under contract numbers 11905117 and 11775024.
%%%%%%%%%%%%%%%%%%%%%%%%%%%%%%%%%%%%%%%%%%%%%%%%%%%%%%%%

%%%%%%%%%%%%%%%%%%%%%%%%%%%%%%%%%%%%%%%%%%%%%%%%%%%%

\end{document}